# A calibration framework for high-resolution hydrological models using a multiresolution and heterogeneous strategy

**Ruochen Sun[1,2,3], Felipe Hernández[1], Xu Liang[1*], and Huiling Yuan[3]**

[1]Department of Civil and Environmental Engineering, University of Pittsburgh, Pittsburgh, PA, USA

[2]State Key Laboratory of Hydrology‐Water Resources and Hydraulic Engineering, College of Hydrology and Water Resources, Hohai University, Nanjing, China

[3]School of Atmospheric Sciences and Key Laboratory of Mesoscale Severe Weather/Ministry of Education, Nanjing University, Nanjing, China

*Corresponding author: Xu Liang (xuliang@pitt.edu)

## Key Points:

- A novel framework is developed for automatic calibration of computationally demanding hydrological models with a large number of parameters.

- The framework alleviates equifinality and calibrated parameters achieve more reasonable values that better correspond to physical reality.

- The framework leads to faster improvement and smoother convergence to optimal objective value and is tested with 134 calibration parameters.

## Abstract

Increasing spatial and temporal resolution of numerical models continues to propel progress in hydrological sciences, but, at the same time, it has strained the ability of modern automatic calibration methods to produce realistic model parameter combinations for these models. This paper presents a new reliable and fast automatic calibration framework to address this issue. In essence, the proposed framework, adopting a divide and conquer strategy, first partitions the parameters into groups of different resolutions based on their sensitivity or importance, in which the most sensitive parameters are prioritized with highest resolution in parameter search space, while the least sensitive ones are explored with the coarsest resolution at beginning. This is followed by an optimization based iterative calibration procedure consisting of a series of sub-tasks or runs. Between consecutive runs, the setup configuration is heterogeneous with parameter search ranges and resolutions varying among groups. At the completion of each sub-task, the parameter ranges within each group are systematically refined from their previously estimated ranges which are initially based on *a priori* information. Parameters attain stable convergence progressively with each run. A comparison of this new calibration framework with a traditional optimization-based approach was performed using a quasi-synthetic double-model setup experiment to calibrate 134 parameters and two well-known distributed hydrological models: the Variable Infiltration Capacity (VIC) model and the Distributed Hydrology Soil Vegetation Model (DHSVM). The results demonstrate statistically that the proposed framework can better mitigate equifinality problem, yields more realistic model parameter estimates, and is computationally more efficient.

## 1. Introduction

Despite of the effectiveness of simple methods for aggregated hydrological estimation (Best et al., 2015; Willems, 2014), the development of complex prediction models (Koster et al., 2017) has enabled a wide array of applications, from large-scale monitoring (Lin et al., 2018; Thielen et al., 2008) to integrated frameworks spanning additional geosciences and engineering branches (Foley et al., 1996; Gochis et al., 2013; Wilkinson et al., 2002). These tools have been made possible due to advances in the ability to simulate land-surface phenomena using physically-based approaches (Clark et al., 2017), together with the growing availability of powerful computational resources (Maxwell, 2013; Tristram et al., 2014). Either because of their fine discretization of space and time (Chaney et al., 2014; David et al., 2011; Koch et al., 2016; Thomas et al., 2017) or because of their massive scale (Souffront Alcantara et al., 2017; Wood et al., 2011), these modern models are characterized by the immense amount of information that they process.

However, the increased complexity of these models requires a fresh contemplation on the fact that almost all hydrological models contain physical and conceptual parameters which cannot be directly measured, such as the porosity and hydraulic conductivity of the soils at different soil depths (DeChant & Moradkhani, 2014), and more so as models become more complex. Uncertainty from these sources have been addressed traditionally by adjusting the model's parameter values manually or automatically such that the simulated response (typically streamflow) matches expected outcomes — usually in the form of available observations (Feyereisen et al., 2007; Peck, 1976; Refsgaard & Storm, 1990; Vansteenkiste et al., 2014). In the manual calibration practice it is well recognized that incorporating expert knowledge or available information could lead to increase in efficiency and accuracy of parameter estimations. However, the manual calibration is of limited value when the number of interacting parameters, parameters whose effects on the response cannot be isolated from one another, is large. Even with a thorough understanding of the physics and mechanisms of the model, the problem becomes intractable when attempting to incorporate the expert knowledge or available information into manual calibration.

Despite of the attempts to systematize manual calibration efforts (Chen & Chau, 2006; Lumb et al., 1994), the advent of complex models necessitates the use of automated, time-efficient methods given the sheer magnitude of the solution space (Abbaspour et al., 2015). Work has been done on various aspects of potential automated solutions. For instance, there is a whole body of literature dedicated just to sensitivity analyses in prioritizing parameters (Cuntz et al., 2015; Pianosi et al., 2016; Song et al., 2015). There also have been major advances in automatic calibration algorithms based on breakthroughs in the field of optimization. Early methods, such as derivative-based algorithms (Gupta & Sorooshian, 1985) or direct search methods — like the Nelder-Mead simplex approach (Nelder & Mead, 1965), utilized local optimization techniques to find locally optimal solutions. Later studies have focused on advanced global optimization methods, which include genetic algorithms (Deb et al., 2002; Goldberg & Holland, 1988; Wang, 1991), differential evolution (Storn & Price, 1997), shuffled complex evolution (Duan et al., 1992, 1994), and particle swarm optimization (Eberhart & Kennedy, 1995; Jiang et al., 2010). In essence, these approaches take advantage of being able to run a model numerous times and use computational algorithms to accelerate the search process.

Optimization based calibration methods, while relatively successful, come with their own set of limitations. Among them is the fact that, when a model possesses many interacting parameters, multiple combinations of value assignments of these model parameters can lead to similar responses — a phenomenon referred to as equifinality of parameter sets (Beven, 1993, 2006; Beven & Freer, 2001). This equifinality phenomenon occurs even for an ideal case, where there are no input data errors, model structural errors or other sources of error, but many of its parameters are interacting with one another. The equifinality phenomenon is especially pronounced: (1) when the number of parameters for calibration is large and the available types of observations (e.g., streamflow, soil moisture, etc.) that can be simultaneously used to minimize the calibration errors are small; and (2) when there are errors in data and in model structures. In other words, when the number of model parameters overwhelms the information content available from observations for calibration, together with various sources of errors, the equifinality could be pronounced. This is especially the case for distributed models. Specifically, many different combinations of the model parameter values can lead to compatible levels of errors in what is known as "attaining compatible errors (e.g., similar Nash–Sutcliffe Efficiency coefficients) for wrong reasons". One of the essential dominant factors of causing such an equifinality phenomenon is the lack of enough constraints and the presence of multiple sources of errors. The equifinality issue related to the model parameter calibration (also called inverse modeling in the groundwater community) is similar to the issue involved in solving a general inverse problem and it is a challenging problem that has not yet been well addressed. An important consequence of the equifinality is that the model results obtained through direct optimization (e.g., simulated streamflow) may be "right for the wrong reasons" (Abbaspour et al., 2007; Götzinger & Bárdossy, 2008; Yang et al., 2008), and therefore potentially ill-suited for prediction purposes. In addition, as it is unable to reliably identify the sources of observed variations from within the parameters, it is incapable of dealing with the inherent uncertainty in parameter identifiability.

Allowing for the creation of stochastic estimates, probabilistic calibration approaches were introduced to mitigate part of this problem. They include the generalized likelihood uncertainty estimation (GLUE) method (Beven & Binley, 2014, 1992), and a wealth of techniques based on Markov Chain Monte Carlo and Bayesian inference (Abbaspour et al., 1997, 2004; Kuczera & Parent, 1998; Laloy & Vrugt, 2012; Vrugt, 2016; Vrugt et al., 2008, 2013; Vrugt, ter Braak, et al., 2009). The problem of equifinality also makes the forecasting applications challenging since the different parameter combinations, together with other sources of errors, can lead to intangible uncertainties in state variables (Carrassi et al., 2018; Moradkhani et al., 2005, 2018). In this respect, various data assimilation algorithms have been designed to account for very large levels of uncertainties and to facilitate estimates that are consistent with both recent observations and previous conditions (Beven & Freer, 2001; Fisher, 2003; Hernández & Liang, 2018, 2019; Ning et al., 2014).

However, modern complex distributed models, containing dozens or hundreds of unknown parameters, pose a major challenge even for the most advanced optimization algorithms. They also demand massive computational resources (Jiang et al., 2010, 2015; Wu et al., 2017). To address these specific challenges, new calibration strategies are needed. Based on our own experiences of using the manual and various automatic calibration methods, we have found that there are certain expert knowledge or available information and some basic principles that could be very useful for calibration if they are used together effectively. We summarize these rules as a set of five principles: (a) determining realistic value ranges for the parameters which are typically expressed in the form of the a priori estimates of the parameter ranges, (b) prioritizing those that are the most impactful and meaningful, (c) establishing cause-and-effect relationships between specific parameters and specific components of the response, (d) addressing the problem sequentially by modifying only one or a few parameters at a time, and (e) narrowing gradually the range of each estimate within a cycle of continuous improvement.

In fact, one or a subset of these aforementioned five principles have been used, but rarely together, by researchers in the past. For example, one effective solution which improves the efficiency of search is to adjust the parameter search space to avoid local minima. Chu et al. (2010) improve the shuffled complex evolution (SCE) method by applying the principal component analysis (PCA) to identify if the parameter search space (i.e., the dimensions of the number of parameters) in SCE is degenerated into a parameter subspace (i.e., with fewer dimensions). If so identified, they restore the searches to the full parameter space. On the contrary, the dynamically dimensioned search (DDS) algorithm (Tolson & Shoemaker, 2007) improves search efficiency by reducing the number of parameters to modify. However, the DSS method does its parameter reduction randomly. That is, DSS blindly selects the parameters to be removed or to be kept for its function evaluations as the process approaches its maximum specified number of function evaluations.

Use of cause-and-effect relationships between parameters and different portions or transformations of the resulting hydrographs has been shown promising (Parada et al., 2003; Sadegh et al., 2016). For example, Parada et al. (2003) present a multiresolution optimization strategy applied to temporal scales to calibrate hydrological models in which the objective functions for calibration are based on the multiresolution framework of wavelet theory. Parada et al. (2003) incorporated their multiresolution approach into the SCE scheme and applied it to calibrate the Three-layer Variable Infiltration Capacity (VIC-3L) model parameters (Liang et al., 1994; 1996a; 1996b; 1999; Liang and Xie, 2001). Their results indicate that the multiresolution paradigm provides more robust calibration than its traditional single-scale counter-part with the SCE scheme. Regarding the multi-objective calibration approaches, there are a number of them (e.g., Boyle et al., 2000; Efstratiadis & Koutsoyiannis, 2008), including a multi-objective version of DDS (Asadzadeh & Tolson, 2013). These efforts have gathered considerable attention (Efstratiadis & Koutsoyiannis, 2010) regarding the constraining of model parameters.

Moreover, the adjustment of parameter value ranges is another effort to improve the calibration (Abbaspour et al., 1997, 2004; Wu et al., 2017). For example, Abbaspour et al. (1997) presented a method called sequential uncertainty domain parameter fitting (SUFI) to calibrate model parameters. The SUFI method first divides each parameter's range into equally sized strata based on the distribution obtained either from an exhaustive stratified sampling scheme or a random stratified sampling scheme. Then, the mean or the midpoint value, depending on the distribution of the parameter, of each stratum is taken to represent the parameter value of that stratum. Next, a hydrologic simulation is run for the all possible combinations of strata for all input parameters (i.e., a brute force approach)--which is the essence of SUFI. The hydrological model simulation results (e.g., soil moisture, pressure head) with all of the parameter combinations are compared with the measurements. The strata that have higher number of hits for each parameter (i.e., the parameter sub-ranges that are more frequently chosen) with smaller differences comparing to the observations (e.g., meet a given criterion or multiple criteria) are kept for next iterations while other strata are removed from future hydrological evaluations. Such iterations continue until a specified number of iterations are reached. The final stratum for each parameter is taken as the solution of the parameter's value or the parameter's range. Abbaspour et al. (1997) show that the SUFI method, applied to models with up to six parameters in their applications, is effective in finding the model parameters' values as the size of each stratum becomes very small (i.e., the method converged) most of the time. The most serious limitation of this SUFI method is that it is not scalable for models with a large number of parameters, e.g., more than 100 parameters. In their study, the largest number of parameters tested was only six.

Following the main idea of Abbaspour et al. (1997), Wu et al. (2017) presented a method to select an optimal parameter range from its initial parameter range for calibration and applied their method to calibrate a rainfall-runoff model (Xinanjiang model) with a total of 10 parameters. The main differences between Abbaspour et al. (1997) and Wu et al. (2017) are that the latter does not iteratively adjust each parameter's range and it is not a brute force approach. The method of Wu et al. (2017) includes three steps: (1) Apply a calibration method, for example, the generic algorithm (GA) method in their study, 100 times to calibrate each of the model parameters. With the 100 calibrated values for each parameter, determine its corresponding probability distribution. (2) For each parameter, determines its new parameter range based on the probability distributions from (1). The new range for the parameter can be (a) corresponding to the minimum range (MINR) or (b) maximum range (MAXR) for a given cumulative frequency (e.g., 50%) or (c) the same as the initial range. Then the optimal range of the parameter is determined by comparing the Nash–Sutcliffe (N-S) efficient corresponding to the calibration results using the new ranges of either (a) or (b) or (c). The choice of the range that leads to the highest N-S efficient is then considered to be the optimal range for the parameter in question. In this selection of the optimal parameter range, it is done in the fashion of varying the parameter range of one parameter at a time while keep other parameters at their initial ranges. (3) Calculate two measures: (a) the sensitivity of the parameter range change to the N-S coefficient for each parameter and, (b) the impact of the range change of one parameter on the calibration of the other parameters through a one-at-a-time fashion comparison. Based on these two measures calculated, decide which parameters' ranges should be changed to their optimal ranges determined in the prior step and which parameters should be kept to their initial ranges. Then calibrate the parameters again using the newly determined ranges with the same calibration method (e.g., GA). The main limitations of the method of Wu et al. (2017) are that: (1) the method does not iteratively adjust the parameter range for each parameter; (2) the method is not scalable for models with a large number of parameters to be calibrated; (3) the improvement with their method of using the selected optimal parameter ranges for each of the 10 parameters measured by the N-S coefficient is negligible (with only 0.01 or less improvement) when comparing to that of using the initial parameter ranges for all of the 10 parameters; and (4) the selection process of the optimal parameter ranges is determined in a pair-wise fashion and thus impacts of the new parameter ranges with multiple parameters at the same time cannot be adequately considered. This probably explains why the utilization of the selected optimal parameter ranges in their study only has improvement of 0.001 in the maximum and of 0.01 in the minimum of the N-S coefficient, respectively.

In this paper we present a new calibration framework with a multiresolution-heterogeneous strategy. The essential idea of our framework is that it not only incorporates, in an automatic manner, all of the five aforementioned principles, but also effectively and seamlessly integrates these principles through a multiresolution-heterogeneous strategy. It is this novel multiresolution-heterogeneous strategy that significantly reduces the search from a very large parameter space to a moderate level; it further enables an effective and efficient search that significantly reduces the equifinality issue among parameters when the number of parameters to be calibrated are very large (e.g., more than 100). In addition, our new calibration framework distinguishes itself from other previous work as in that the existing methods simply incorporated one or a subset of the five aforementioned principles (i.e., used in isolation), as briefly reviewed above. Notably also is that their implementations of some of the principles are different from what we present in this study even though the same terminology is shared, such as "reducing the number of parameters to calibrate at a time". More specifically, our framework partitions a calibration problem into a sequence of optimization sub-tasks. Each sub-task is associated with parameter groups of different resolutions and different parameter ranges in the parameter search space according to the importance of parameters. Parameters that are sensitive or important are ranked the first group, and the least important ones the last group. The search ranges of the parameters are iteratively determined and dynamically updated between consecutive sub-tasks. In other words, across sub-tasks, the manner the resolution is set is heterogeneous. The structure and the construction of the framework proposed in this study make it possible to reduce uncertainty where it matters the most, to attain more realistic estimates, and to increase the overall computational efficiency of the entire process. This paper is organized as follows. The methodology is described in detail in Section 2, the setup of a testing case study using double-model setup is explained in Section 3. The results comparing our new framework with a traditional automatic approach are presented in Section 4 for a synthetic experiment study. Conclusions are provided in Section 5.

## 2. Methodology

A hydrological model can be represented by a function $M$ which produces a vector of outputs $\boldsymbol{o}^{t+1}$ and a vector of state variables $\boldsymbol{s}^{t+1}$ at time $t + 1$ from a vector of state variables $\boldsymbol{s}^t$ at time $t$, a vector of forcing values $\boldsymbol{f}^{t+1}$ at time $t + 1$, and a vector of static parameters $\boldsymbol{p}$:

$$\boldsymbol{o}^{t+1}, \boldsymbol{s}^{t+1} = M(\boldsymbol{s}^t, \boldsymbol{f}^{t+1}, \boldsymbol{p}) \tag{1}$$

The outputs $\boldsymbol{o}$ correspond to fluxes such as the streamflow at each of the channels and the evapotranspiration from each sub-watershed (or cell/pixel if the model is gridded). The state variables $\boldsymbol{s}$ include, for example, the soil moisture and temperature, and the depth of the snow pack. Forcing $\boldsymbol{f}$ include precipitation, air temperature, air pressure, wind, humidity, and solar radiation. The parameters $\boldsymbol{p}$ include the soil hydraulic conductivity and porosity, the percentage of impervious land, and the friction factor of the channels. The model can also be run as an extended period simulation between an initial time $t_i$ and a final time $t_f$ by iteratively using Eq. (1):

$$\boldsymbol{o}^{t_{i+1}:t_f} = M(\boldsymbol{s}^{t_i}, \boldsymbol{f}^{t_{i+1}:t_f}, \boldsymbol{p}) \tag{2}$$

For the results of the model $\boldsymbol{o}^{t_{i+1}:t_f}$ to be agreeable with reality, $\boldsymbol{s}^{t_i}$, $\boldsymbol{f}^{t_{i+1}:t_f}$ and $\boldsymbol{p}$ need to be properly estimated. Calibrating model $M$ can be defined as the process of finding a set of parameters $\boldsymbol{p}_*$ such that the model's outputs match a set of observations $\boldsymbol{o}_{\text{obs}}^{t_{i+1}:t_f}$ as close as possible, given known values of $\boldsymbol{s}^{t_i}$ and $\boldsymbol{f}^{t_{i+1}:t_f}$. That is, the calibration seeks to find $\boldsymbol{p}_*$ such that one or more error metrics $\boldsymbol{e}\left(\boldsymbol{o}^{t_{i+1}:t_f}, \boldsymbol{o}_{\text{obs}}^{t_{i+1}:t_f}\right)$ are minimized (note that if more than one error metric is used, $\boldsymbol{p}_*$ is usually non-unique):

$$\boldsymbol{p}_* = \operatorname*{argmin}_{\boldsymbol{p}} \boldsymbol{e}\left[M(\boldsymbol{s}^{t_i}, \boldsymbol{f}^{t_{i+1}:t_f}, \boldsymbol{p}), \boldsymbol{o}_{\text{obs}}^{t_{i+1}:t_f}\right] \tag{3}$$

The proposed calibration framework, aimed at solving the problem posed in Eq. (3) for vectors $\boldsymbol{p}$ of high dimensionality, consists of two phases: a strategy phase and an iterative optimization phase. The strategy phase controls how the calibration will be conducted in the optimization phase by determining the reasonable variation range for each of the parameters, ranking their importance, deciding in what order they should be estimated, and selecting the number and length of the iterations to be performed. This phase involves a modest amount of expertise and modeling judgement from the user, which could be streamlined with proper guidelines. The optimization phase can be fully automated and runs a population-based optimization algorithm iteratively. In each iteration, the focus is set on a different subset of the parameters. These iterations progressively reduce the uncertainty on the estimates for $\boldsymbol{p}$ in a multiresolution and heterogeneous manner.

Figure 1 shows the flow chart of the proposed calibration framework, with the strategy phase consisting of eight steps on the left and the optimization phase consisting of steps on the right. The algorithm has four main meta-parameters: the number of parameter groups, $g$; the multiplier, $w$, which dictates how the standard deviation is scaled to establish an updated variation range of parameters; a function to compute the number of evaluations, $n(r)$, given the run number index, $r$; and the number of final fine-tuning runs, $x$. The value $x = r - g$ defines the extra number of fine-tuning runs to be carried out. Each extra fine-tuning run allows all the parameters to be modified simultaneously at their finest resolution level where each parameter range has been reduced to the narrowest. Here we detail each of the steps of the framework and the purpose of these meta-parameters.

## 2.1. The strategy phase

The first four steps of the strategy phase, as listed in the left column of Figure 1, seek to identify the relative importance of each of the parameters of the model. Some parameters that are deemed unimportant based on experience are assigned constant default values in Eq. (2) and will not be subjected to calibration (step 1). For the rest sensitive parameters, a sensitivity analysis is performed (step 2) in which different values are assigned to each parameter to assess the effect on the error metrics $\boldsymbol{e}$ after running the model Eq. (2). The selection of the sensitivity evaluation methods (e.g., Pianosi et al., 2016; Song et al., 2015) requires that the analysis is capable of providing sufficient information for distinguishing the importance or relevance of sensitive parameters (step 3). After the sensitivity analysis, the parameters that are classified as sensitive are sorted into $g$ groups, with the first group consisting of the most relevant or sensitive parameters, and the last group the least relevant (step 4). The additionally identified non-sensitive parameters are also assigned constant values.

In selecting the number of parameter groups, $g$, we offer the following rule of thumb: (1) We recommend to keep the number of parameters in each group relatively balanced with each other, i.e., with each group containing roughly equal number of parameters rather than having the number of parameters in one group, for example, significantly larger than other groups. (2) The number $g$ itself may be affected by the degree of automation. If the optimization phase is not fully automated, larger values of $g$ will result in additional processing steps in-between runs of the optimization algorithm. Fully automated algorithm does not have this overhead, and thus may select a larger $g$ for a more detailed analysis. However, we do not recommend separating equally-relevant parameters into different groups for the sake of increasing $g$, as this may translate into some of them being disproportionally emphasized or de-emphasized in the search.

Step 5 determines the initial widest range of variation $\left[p_{i,\min}, p_{i,\max}\right]$ for each of the sensitive parameters, $p_i$. These ranges should be bounded by realistic values and, may be acquired from empirical or expert knowledge. Making use of empirical tables for standard soil textures (Meyer et al., 1997) would be one such example. In some cases, using an exponential transformation to the variation range of specific parameters can be advantageous (e.g., the permeability of soils is spread more uniformly in the conductivity range when using an exponential scale).

The ranges of search do not have to be equal to the ranges of variation of the parameters. Parameters values are searched with different granularities or resolutions at different point throughout the entire process: some are searched within their full ranges, others are only selected from discrete interior points. Based on their priority and hierarchy determined in step 4, the present framework applies different granularities or resolutions in the search (i.e., performs a multiresolution search) and uses a heterogeneous treatment in each range. This facilitates an efficient and robust optimization by searching a wider range in the targeted parameter groups, but only discrete points for non-focused groups.

Step 6 determines the discrete point selections for each coarse resolution based on parameters' variation ranges. For example, if one parameter (e.g., par1) has a range of [0, 1], a coarse or low resolution search may be {0.3, 0.6}, an intermediate resolution search {0.2, 0.5, 0.9}, while a fine or high resolution search may be the whole range of [0, 1]. When par1 is the main target of calibration, this high resolution range is adopted; otherwise, its discrete coarse resolution is adopted. Such a coarse resolution search for non-targeted parameters provides some degree of variation for par1, but does not drastically increase the complexity of the search since it avoids a whole range search. Thus, the proposed multiresolution and heterogeneous calibration strategy addresses the critical issue of "When and what resolution should be used for searching for each parameter?" In contrast, traditional optimization-based calibration methods, rather than calibrate in a hierarchical manner, treat all parameters equally without partiality. In other words, they use the same resolution and consider its full range for each parameter. As a result, the traditional calibration conducts search in an immense solution space and is more vulnerable to equifinality problems, especially when strong parameter interactions are present.

After the first six steps are completed, a running or execution plan for the optimization phase is established. The running plan (steps 7 and 8) determines the total number of runs, the target or focused groups of parameters for each run, and the number of candidate solutions or parameterizations to evaluate in each run (i.e., how many times Eq. 2 is run). That is, the proposed execution plan divides the entire calibration problem into a series of sequence of runs or sub-tasks with heterogeneous treatments across runs. Within each "run", all the parameters are adjusted within their corresponding ranges to minimize the objective functions which is represented by the error metrics $e$. After each run, the ranges of the parameters are updated and the focused parameter group is shifted in the subsequent run in order to prioritize the most relevant ones, and the resolution used is accordingly changed across runs. Details are explained below in the optimization phase.

## 2.2. The optimization phase

Depending on the configurations set up in the strategy phase, the automatic calibration framework allows for some procedural variations from its fundamental construct, which is presented in the right column of Figure 1. During the first run, parameters of the first group are the optimization focus. These focused parameters are searched in the high resolution by having their full ranges explored, while parameters in the other groups are only searched in coarser resolutions among discrete values. For the second run, the second parameter group becomes the target and is searched in the high resolution over their full ranges. The first group is still searched in the high resolution in the second run but over reduced variation ranges based on the previous run results, while parameters in the other lesser important groups are again searched in the coarser resolutions over discrete points. In each subsequent run, the previously targeted groups are continually searched in the high resolution but using reduced or shrunk parameter ranges, the current focus group is searched in a complete range, and lower priority groups over discrete points.

Figure 2 illustrates this procedure conceptually with an example case involving three parameters associated with three groups and a running plan of four runs, in which the parameter 1 (Group 1) is the most important or the highest priority one, and parameter 3 (Group 3) the least important one. Table 1 provides details of a test case used in which parameters are classified into six groups and the running plan includes a total of seven runs. Each run also employs a different number of model evaluations. For the first run, the model is evaluated 200 times, i.e., 200 different parameter combinations are evaluated. With Group 1 being the focus, parameters in Group 1 have a full range search with a high resolution, while parameters in Groups 2 to 6 are searched among discrete points. After the first run, the parameter ranges in Group 1 are reduced or "shrunk" as denoted in Table 1. For the second run, Group 2 is the focus and the parameters are searched in their full ranges, and those in other groups continue to be searched on discrete points, except for Group 1 which is searched in shrunk ranges with the high resolution. That is, parameters in both Groups 1 and 2 are now searched at fine resolutions while parameters in all other groups are searched at coarser resolutions. This procedure is repeated. At the completion of Run 6, an additional fine-tune run (i.e., Run 7) is added allowing all of the parameters be searched at fine resolutions.

The optimization configuration presented reflects the guiding principles underlying the design of the proposed calibration framework: (a) prioritization of sensitive parameters; (b) a divide-and-conquer approach to constraint the parameter search range and resolution; and (c) a progressive reduction in the variability of the estimated parameters. The core ideas in the new calibration framework manifest a robust multiresolution and heterogeneous calibration strategy.

The optimization phase is carried out using a set of assignments to $\boldsymbol{p}$ which evolves from one run to the next. This set is called a population per the terminology in the field of the evolutionary computation. The set is initialized randomly by taking into account the parameter ranges. If the group is designated for a full range search, values are sampled using a uniform distribution; if it is put forth for a discrete search, only the discrete values are randomly selected. After the first run, if a parameter has already been fully explored, then a shrunk variation range is used for subsequent runs based on the evolved solution set. From the current (evolved) population, the mean $\mu$ and standard deviation $\sigma$ for each parameter are determined. The new range for a parameter $p_i$ is determined using a meta-parameter $w$ as follows:

$$\max(p_{i,\min}, \mu_i - w\sigma_i) \leq p_i \leq \min(p_{i,\max}, \mu_i + w\sigma_i) \qquad (4)$$

Large values of $w$ allow for a more gradual reduction of the parameter ranges from one run to the next. Aside from the first run (i.e., Run 1), the optimization algorithm does not start from a completely random initial population, but from one obtained that has already achieved a certain level of convergence towards the objective functions. However, allowing part of one run's initial population to be determined randomly should help in balancing the inherent conflict between exploitation and exploration ( Chen et al., 2009) by increasing the solution diversity (Leung et al., 1997).

Each evaluation involves computing the objective functions, or error metrics, $\boldsymbol{e}$. For the run number index, $r$, $n(r)$ number of evaluations are carried out; this $n(r)$ serves as a control parameter and is up to the user to determine it. In the example shown in Table 1, $n(r)$ increases with $r$. The reason behind this adoption is to allow for more explorations as $r$ increases to deal with the growing number of parameter groups that are searched in high resolution. In other words, as more parameters with high resolutions are added to be searched, more model evaluations should be allocated in order to accommodate the increased number of parameters to be searched with high resolutions while the number of parameters stay the same in each run (see Table 1 and Figure 2). That is, increasing the model evaluations in each run would let the parameters in latter groups receive sufficient chances to be searched. In addition, since the parameters associated with better ranks are more sensitive, a fewer number of model evaluations would be enough to effectively shrink these parameters' ranges (e.g., those in the first group). A test of this hypothesis -- increasing model evaluations -- is discussed in subsection 3.2. In general, one should experiment with $n(r)$ selection for specific applications. In the case involving a single objective function, $e$, the solution is the parameter set that performs best in terms of the objective. If multiple objectives are involved, this solution may take the form of a Pareto Front (Deb, 2014), which includes solutions that performed well on each of the individual metrics and also those that offer the best trade-offs among them.

In summary, two important ideas are involved in our new calibration framework: a multiresolution search and heterogeneous treatments. Implementation of these two ideas allows important parameters to be explored more thoroughly in the parameter search space with higher resolutions or granularities than those in the less important groups which are only allowed to vary at a much coarser resolution at the earlier stages. The variation ranges of all the parameters are systematically and iteratively adjusted from one run to the next. However, the less important parameters are allowed to vary at fine resolutions at a later run when the parameter ranges in the more important groups are reduced as shown in Figure 2 and Table 1. The calibration strategy is, therefore, hierarchically adjusted. This is why the new calibration framework so designed can reduce equifinality and accelerate the whole calibration process by reducing the size of the search space through multiresolution in each run.

### 2.3. The optimization algorithm

Our calibration framework can utilize, in principle, any population-based optimization algorithm; that is, algorithms that are given an initial population of parameter solutions, $\boldsymbol{p}$, can return with an optimized population. If multiple error metrics $\boldsymbol{e}$ are defined, the optimizer needs to support multiple objectives or, alternatively, the various metrics need to be lumped into a single formula. In the literature there exist many evolutionary multi-objective algorithms that could be used (Deb, 2014). For the proposed framework, the optimizers need to be able to encode candidate solutions using both continuous and discrete variables to carry out the multiresolution treatment. This requirement is not common for optimization problems and, thus, there are not many such algorithms directly available — usually optimizers focus on "global" optimization problems (for continuous variables) or on "combinatorial" optimization problems (for discrete variables) (Papadimitriou & Steiglitz, 1998).

In our implementation of the new calibration framework we used an optimization algorithm (Hernández & Liang, 2018) which does allow for the representation of continuous and discrete candidate solutions simultaneously. Our algorithm (available at https://github.com/felherc/MAESTRO_MO) follows a recent and successful approach which consists of making use of an ensemble of cooperating metaheuristics (Peng et al., 2010; Y. Wang et al., 2013). The simultaneous use of multiple low-level search heuristics allows one to mitigate problems associated with the "no free lunch" theorems (Droste et al., 2002; Wolpert & Macready, 1997): the inability of a single optimization strategy to perform consistently well on problems of different nature. In other words, since it has been proven that it is impossible for a single optimization strategy to have a superior performance in all types of problems, this ensemble approach helps diversify the pool of available strategies aiming to increase performance in general. The algorithm is similar to AMALGAM (Vrugt, Robinson, et al., 2009; Vrugt & Robinson, 2007), featuring an adaptive technique to determine how intensively to query each of the low-level optimizers in the ensemble and the ability to run evaluations in parallel. The optimizer has been previously used successfully for high-resolution hydrologic data assimilation (Hernández & Liang, 2018). Two low-level metaheuristics were used within the ensemble: the established NSGA-II (Deb et al., 2002) and a hybrid between a Metropolis-Hastings sampler (Haario et al., 2001) and Ant Colony Optimization for discrete (Dorigo et al., 2006) and continuous (Socha & Dorigo, 2008) variables.

## 3. Experimental design for a synthetic case study

To assess the merits and performance of the proposed framework, we carry out a comparison study based on a synthetic case study because the 'true' parameter values are known. This allows for an in-depth and meaningful comparison between the calibrated and the true parameters. Basically, we use as the measures the accuracy of the estimated parameter values in comparison with the known true values, not just based on the goodness of the fit between model simulated variables and observations (e.g., streamflow). Here, we are interested in determining the quality of the parameter estimates rather than the weaker model's goodness of fit to observations. As the latter has a pitfall for it operates under the consideration that a better fit of observations implies parameter estimates are closer to true values and thus gives better predictability. But it could fail because of the likelihood of "good fit for the wrong reasons" as previously elaboration on equifinality.  To avoid dealing with the intangible issue of decomposing the individual sources of errors when large unknown errors exist (Renard et al., 2010), our study focuses on equifinality caused by the effectiveness of the calibration approach and model structure errors, and not those caused by errors/uncertainties involved in the model input data and observations.

This section first presents how we designed a complex high-dimensional synthetic calibration problem to compare the performance of the new calibration framework with that of a traditional evolutionary optimization-based calibration approach. Then we introduce three different configurations using the new framework and, finally, we define the performance metrics that were used to evaluate them.

### 3.1. Synthetic case design and construct

Real world observational data, if not extensive or not properly distributed, are not adequate for testing and evaluating the performance of our new calibration framework. This lack of sufficient real world observation for assessment purpose prompts us to opt for a common approach (Evin et al., 2014; Schoups & Vrugt, 2010) in which results from a new calibration methodology are benchmarked against known synthetic model results. The synthetic model, called the *reference model* hereafter, in which the parameters are known, is used to generate a time series of outputs. These outputs then serve as the observations. The *initial model* is the model used to obtain the calibrated parameters based on the observations. The calibrated model parameters obtained from our presented framework are then compared to those known parameters used for the reference model. The main requirements for the double-model setup are that the two models should have different model structures, and that they share sufficient number of common parameters. In order to incorporate model structural errors, which are prevalent when simulating natural processes, the reference model and the initial model used in this study are represented by two different modeling engines with substantially different modeling structures. This makes our synthetic testing approach significantly different from others in previous studies that used only one model engine. The use of two different modeling engines, i.e., double-model setup, allows us to test whether the presented framework has the capability to overcome the model structure uncertainty and yield good parameter estimates (Renard et al., 2010).

We used two popular distributed hydrological modeling engines, namely the Variable Infiltration Capacity (VIC) model (Liang et al., 1994; 1996a; 1996b; 1999; Liang & Xie, 2001) and the Distributed Hydrology Soil and Vegetation Model (DHSVM) (Wigmosta et al., 1994, 2002) to construct the synthetic experiment. VIC was originally designed to model large watersheds by taking into account of the effects of spatial subgrid variability of precipitation, soil properties and vegetation cover on soil moisture and surface fluxes (e.g., evapotranspiration). The DHSVM, on the other hand, was developed to numerically represent the effects of local weather, topography, soil type, and vegetation on hydrological processes within relatively small watersheds using high spatial resolution. In this study the VIC modeling engine is used for the reference model and the DHSVM for the initial model. We use the French Creek watershed (Figure 3) in Pennsylvania for this study. This watershed has a drainage area of about 160 km$^2$, which is suitable for both models. VIC was configured at a 1/32° resolution with a daily time step, while the DHSVM ran at a 500-m resolution with an hourly time step. The layout of the VIC modelling grid for this watershed is also shown in Figure 3. The daily discharge records of the basin outlet from years of 2003-2011 were collected from U.S. Geological Survey (USGS) station 01472157. When contrasting the simulated discharge, the time series produced by the DHSVM was aggregated to daily discharge. The meteorological forcing data used to run both models were collected from Phase 2 of the North American Land Data Assimilation System (NLDAS-2) forcing dataset (Cosgrove et al., 2003).

We first used the VIC model to generate the synthetic daily "observed" streamflow $o_{obs}^{t_{i+1}:t_f}$ from 2003 to 2011, and then we calibrated the DHSVM model parameters against the synthetic streamflow from VIC to maximize the Nash–Sutcliffe Efficiency (NSE) coefficient (Nash & Sutcliffe, 1970) which is expressed as,

$$NSE = 1 - \frac{\sum_{i=1}^{n}(S_i - O_i)^2}{\sum_{i=1}^{n}(O_i - \overline{O})^2}$$

(5)

where $n$ is the total number of time steps, $O_i$ and $S_i$ are the daily observed and simulated streamflow at time $i$, and $\overline{O}$ is the observed mean value. The calibrated parameters that are common between VIC and DHSVM are the soil's field capacity, porosity, wilting point, vertical saturated hydraulic conductivity, and the vegetation's minimum stomatal resistance and radiation attenuation. Each of these six parameters has an actual physical meaning and plays the same role in these two models. Therefore, the values of these parameters defined in the VIC model are used as reference "true" values. In addition to these physically-based parameters, there are six more conceptual parameters (Table 2) in the VIC model which can have impacts on simulated streamflow. Therefore, for the sake of realism, we first calibrated the VIC model with the real observed streamflow from the USGS station and the two soil types from the CONUS-SOIL dataset (Miller & White, 1998) in setting these six VIC model parameters. The calibrated values of the six conceptual parameters are shown in Table 2, which resulted in an NSE of 0.6. Figure 4 shows the simulated hydrograph using the calibrated VIC parameters and the USGS observed hydrograph along with the areal precipitation time series. Although the NSE is not very high and the time series presents scattered underestimations of peak flows, the simulated hydrograph has the same trend as the observed hydrograph and the areal precipitation time series with good depiction of the baseflow. Therefore, the VIC model can be viewed as a reasonably reliable tool to generate the synthetic "observed" streamflow in this watershed.

The French Creek watershed has only two soil types according to the CONUS-SOIL dataset: silt loam (92%) and loam (8%). To increase the dimensionality of the parameter space (given that DHSVM assigns parameters per soil type), and thus the complexity of the calibration problem, we randomly assigned each VIC model cell a unique soil texture, for a total of 21 textures. The textures were selected from the soil texture triangle defined by the U.S. Department of Agriculture (USDA) (Ditzler et al., 2017). The white dots in Figure 5a show their locations in the triangle, which were selected for maximum heterogeneity. The corresponding soil parameter values for each VIC cell were identified based on the percentages of sand and clay (Saxton et al., 1986; Saxton & Rawls, 2006). Regarding land cover, this watershed is dominated by deciduous forests (47%). We set the vegetation parameters to the default values based on the corresponding vegetation types. By using the six calibrated conceptual parameters of VIC with the 21 synthetic soil textures, we obtained the synthetic "observed" streamflow from running the VIC model.

We then set up the DHSVM model with the same soil textures and land cover types. The DHSVM model cells that are located within one VIC cell were assigned the corresponding soil texture of that VIC cell. Figure 5b shows the soil type distribution for the DHSVM model. The ID numbers correspond to the 21 soil textures which are assigned to the 21 VIC cells. We conducted a sensitivity analysis using 2-level fractional factorial experiments (Liang & Guo, 2003; Montgomery, 2012) in determining which parameters to include in the calibration process. During the analysis, a single soil type and a single vegetation type were assumed for the entire watershed. That is, the sensitivity analysis is not carried out at the granularity of each soil and vegetation type for each parameter. For each of the candidate soil and vegetation parameters, a "low" and a "high" assignment was defined. A total of 128 model evaluations were used for this sensitivity analysis. Although more comprehensive sensitivity analysis methods exist (Cuntz et al., 2015; Matthias et al., 2015; Pianosi et al., 2016; Song et al., 2015), this factorial approach is sufficient for our purpose which is to qualitatively identify the model parameters' importance and group them according to their sensitivities. From this perspective, a simpler and faster sensitivity analysis method is desirable and the results obtained agree with those from the perturbation-response method by Du et al. (2014) as detailed later.

The sensitivity analysis helped to select six sensitive parameters to be calibrated for each soil type in the DHSVM model. These six parameters include the four soil-related parameters that also exist in the VIC model, and the exponential decrease factor for the vertical hydraulic conductivity (i.e., the $5^{th}$ parameter) and the lateral saturated hydraulic conductivity (i.e., the $6^{th}$ parameter). As for the vegetation parameters, we only calibrated some of the dominating vegetation type (deciduous forest), i.e., the moisture threshold for the over-story and under-story, the aerodynamic attenuation, and the fractional coverage. Along with the general Manning's coefficient of the river channels, a total of 134 parameters are calibrated in the DHSVM model. Among them, 87 parameters are also in the VIC model and they are used as the reference "true" values. Table 3 provides a detailed summary of the common soil parameters corresponding to the 21 soil textures. The soil parameter ranges of different soil texture types in Table 3 are provided by Meyer et al. (1997). The rest of the parameters which need to be calibrated in the DHSVM model are shown in Table 4. Their corresponding ranges are based either on typical ranges applied in former studies or on their physical constraints. Given that the vertical saturated hydraulic conductivity is the least relevant parameter as determined through the sensitivity analysis, and its parameter ranges span several orders of magnitude, we used a base-10 logarithmic scale for its range during the calibration process.

Because of the stochastic nature of these calibration approaches, the relative performance of different calibration schemes must be assessed over multiple optimization trials. For this study we carried out ten trials. Each trial is identical to one another in the experiment setup and the initial value used for each parameter is randomly generated for each parameter based on its specified range. Each trial evokes 4,000 times of hydrological model evaluations.

## 3.2. Calibration schemes

For the strategy phase of the calibration, we first used the information gathered from the 2-level fractional factorial sensitivity analysis (i.e., the p-values for individual parameters and for multi-parameter interactions) in determining the relative sensitivity rankings of the 134 parameters. Our results (not shown) were consistent with the conclusions of Du et al. (2014) regarding parameter sensitivity of the DHSVM model, where Du et al. (2014) applied a local sensitivity analysis method -- a perturbation-response method -- which explores changes of model response by varying one parameter at a time. We then decided to divide the 134 parameters of the DHSVM model into $g = 6$ groups, with each group containing roughly equal number of parameters. The most sensitive parameters are assigned to the first groups so that they get more chances to be adjusted. Table 5 lists the grouping of parameters in our study. As can be seen, each of the top three groups contains field capacity, lateral saturated hydraulic conductivity and the exponential decrease factor of seven types of soil texture, which correspond to seven VIC model cells/pixels. Those three soil parameters are all very sensitive, so we were reluctant to rank them further in order of priority. Since no sensitivity analysis was carried out at the granularity of each soil type for each parameter, these soil property-related parameters should then, in principle, be kept in the same group based on our guidelines from the sensitivity analysis. As there are 21 different soil textures in the case study, this would lead to over 60 parameters out of 134 parameters to be placed in group 1, which would make the number of parameters among the six groups highly unbalanced. In fact, such kind of unbalanced situation could occur often, especially for distributed models. To mitigate this unbalanced grouping problem, we divided the 21 soil types into three groups, i.e., groups $1 - 3$. Each group included seven soil types and each soil type included three soil parameters which are the field capacity, lateral saturated hydraulic conductivity, and exponential decrease component (see Table 5). Because we did not have any prior knowledge regarding which soil types are more sensitive than others to be placed in group 1, we have investigated three configurations. That is, we put them into the top three groups and divide them by their cells in the VIC model. The problem of determining which of these cells to give priority is thus further investigated in this study. In particular, we tested three calibration configurations corresponding to three ways to divide these cells. The three configurations were compared with the traditional calibration scheme, in which all of the 134 parameters selected for calibration are simultaneously estimated with the same resolution over their ranges defined in Tables 3 and 4.

It is worth mentioning that it is sometimes impractical and counterproductive even to carry out a detailed sensitivity analysis at the granularity of each soil type, for example, a sensitivity analysis for the 21 different soil textures in this case. This is because the sensitivity results could hardly provide helpful information to distinguish their sensitivities from one another due to their narrow and similar variation ranges, such as the ones for field capacities (see Table 3). In such situations, we recommend users to apply complementary criteria based on modeler's expertise or to explore different configurations as we illustrate here in this study in determining the ranks of these groups. In fact, to what granularity a sensitivity analysis should be performed and what strategy it should be taken to complement it is a problem that is common to distributed models but has no consensus at this time. It is one of the important issues to be further investigated.

The three configurations are constructed by dividing the watershed into three regions: downstream, midstream, and upstream, based on the flow time to the outlet (Figure 3). The three configurations and the traditional calibration approach, which is used as a control, are described as follows:

1. *ExpD-U*: The soil parameters corresponding to the seven VIC cells of the downstream region are in the first group, followed by midstream cells and upstream cells in the second and third group, respectively. For the fourth group, they include porosity and wilting point for ten types of soil texture that correspond to the ten cells with top ten shortest flow time, including seven cells of the downstream region and three cells of the midstream region. The soil parameters corresponding to the remaining 11 cells are in the fifth group.

2. *ExpU-D*: Instead of calibrating parameters from downstream to upstream, ExpU-D reverses the calibration order of ExpD-U. That is, the soil parameters corresponding to the seven VIC cells of the upstream region are in the first group, and the soil parameters for the downstream cells are in the third group. Similarly, the soil parameters for the ten cells with top ten longest flow times are in the fourth group. The fifth group consists of soil parameters corresponding to the remaining 11 cells.

3. *ExpRand*: Since the VIC cells are divided based on the ID numbers of soil texture in the DHSVM model (Figure 5b), and each soil texture corresponds to only one VIC cell, the ID numbers also represent VIC cells. Parameters for cells with ID 1-7 are in the first group, and so on. For porosity and wilting point, cells with ID 1-10 are in the fourth group, and cells with ID 11-21 are in the fifth group.

4. *ExpTrad*: All 134 parameters are calibrated together. This scheme represents traditional automatic calibration approaches and is therefore used as a control. It is important to note that the same sensitivity analysis used to determine the parameter importance (i.e., parameter groups) for the new framework was also the one used to determine which parameters to calibrate for ExpTrad. Thus, both ExpTrad and our new framework (i.e., ExpD-U, ExpU-D, and ExpRand) had the benefit of the same sensitivity analysis process and had the same 128 model evaluations.

The setup for the three configurations using the new framework is presented in Table 1. With the selected $n(r)$ for each run, the total number of model evaluations summed up to 4000. The selected evolution of parameter ranges and the number of discrete values for each group are also given in Table 1. Regarding the selected $n(r)$ for each run $(r)$, their number should increase with each run, but do not need to be in the specific numbers as we listed in Table 1, since one can select different numbers based on the individual models used and the computational resources available. The reason we used $n(r)$ to increase with $r$ is to follow our rationale discussed in Section 2.2. To test that hypothesis, we investigated the impacts of the allocation strategy to the number of model evaluations, i.e., $n(r)$, in each run. We tested three setups for $n(r)$: (1) to increase with $r$, (2) to decrease with $r$, and (3) to remain constant. We found that, confirming our hypothesis, 'to increase with $r$' is most effective. However, this hypothesis may still need to be more thoroughly explored in the future.

### 3.3. Evaluation metrics

We developed a set of metrics to evaluate the performance of the different configurations based on the new framework (i.e., ExpD-U, ExpU-D, and ExpRand) and the performance of ExpTrad based on the 20 best solutions/parameter combinations produced for each trial. In particular, the Tukey boxplot (Montgomery, 2012) of each parameter was computed from the 20 optimized results of a trial. These boxplots were used in the selection of the performance metrics. The bottom and top of the box correspond to the first (Q1) and third (Q3) quartiles, which are the 25th and 75th percentiles of the 20 solutions, respectively. The interquartile range (IQR), which is a measure of statistical dispersion, is equal to the difference between the third and first quartiles (IQR = Q3 - Q1). For the Tukey boxplots, the lower whisker represents the lowest solution still within 1.5 IQR of the lower quartile (Q1 - 1.5×IQR), and the upper whisker represents the highest solution still within 1.5 IQR of the upper quartile (Q3 + 1.5×IQR). These Tukey boxplots for each parameter and each configuration (i.e., ExpD-U, ExpU-D, ExpRand, and ExpTrad) can be used to measure how widespread the best solutions for each configuration are. It is clear that the narrower the spread the better; and also, the closer the estimates to the true values the better. In addition, the fewer the number of parameters hit the boundary the better as hitting the boundary also casts doubt on the physical validity of the resulting estimates. Since there are too many Tukey boxplots to include, we developed a set of metrics to evaluate the performance based on these Tukey boxplots. These statistic metrics are defined as follows:

1. *No convergence* (NC): If the width between the two whiskers is greater than 10% of the original parameter range, it is considered that the parameter reached no convergence. *NC* is the total number of such parameters for a solution set. The non-convergence of parameters stems from equifinality and can mean that the algorithm was unable to find satisfactory assignments that perform consistently well. In this sense, a large NC number can be seen as an indication of poor performance and, moreover, a higher level of equifinality.

2. *Hitting the boundary* (HB)*:* If the width between the median and the boundary of the parameter range is less than 1% of the original range, or the whiskers of the box reach the boundary, this parameter is considered to have hit the boundary. Realistic values of parameters are rarely located at the extremes of the variation intervals and, more often, convergence to the boundary usually means that the algorithm was seeking unrealistic values to compensate for failures in the assignments to other parameters. Therefore, a large HB number also indicates poor performance. Parameter estimates hitting the boundary are also a hallmark indication of equifinality problems, where assignments are pushed to be unreasonably low or high to compensate for poor combinations with other parameters' estimated values.

3. Ratio of smaller *Absolute relative error* (RosARE)*:* This metric is a relative score between the results of a configuration with the new framework and the traditional method. Among the 87 parameters with referenced true values, the parameters that have converged in both the new framework-based configurations and the traditional calibration approach are selected. The absolute relative error (ARE) of each of these parameters is calculated using the referenced true values. The RosARE is then computed as the number of parameters with smaller ARE in the new framework-based configuration divided by the total number of parameters that converged. If the RosARE is higher than 50%, the new framework-based configuration yielded more calibrated parameters that approach their "true" values than the traditional calibration approach.

## 4. Results and discussion for a synthetic case study

Because of the stochastic nature of these calibration approaches, the relative performance of different calibration schemes must be assessed over multiple optimization trials. We have conducted ten trials. We performed a one-way Analysis of Variance (ANOVA) for the NC and HB scores, and a 1-Sample t-test for the RosARE to determine the statistical significance of the difference between the three configurations based on the new calibration framework and the traditional approach (i.e., ExpTrad). We found that all of the three configurations with the new framework produced statistically significantly fewer numbers of NC and HB than the traditional calibration approach. It can be seen in Figure 6 that the Tukey's 95% confidence intervals of differences of means for NC and HB between the three configurations (i.e., ExpD-U, ExpU-D, and ExpRand) and the traditional approach (ExpTrad) do not contain zero, indicating that the corresponding means are significantly different. The p-values of the two hypothesis tests for NC and HB are 0.004 and 0.006, which are smaller than the defined significance level of 0.05. In the 1-Sample t-test for RosARE, we assumed the hypothesized mean was 50% and set the significance level to 0.05. For the ten trials, only ExpD-U produced a mean RosARE that was statistically significantly higher than 50% with p-value = 0.016. Although the other two configurations, ExpU-D and ExpRand, produced a mean of RosARE higher than 50%, their p-values were greater than 0.05.

The means of the three evaluation metrics of the ten trials are listed in Table 6. The three configurations based on the new framework have significantly better performance than ExpTrad in terms of NC and HB. The number of not converged parameters for ExpTrad is about five times (for ExpD-U) or nearly four times (for ExpU-D and ExpRand) of those of the three configurations with the new framework. It can be seen in Table 7 that most of the parameters that did not converge with the new framework are in the last group, which had the lowest chances to be adjusted. However, the parameters that do not converge for ExpTrad are distributed throughout all groups, which clearly shows the impacts of parameter interactions in the parameter search space. No convergence in the parameters is a serious problem and puts a calibration effort in jeopardy. When parameters hit the upper or lower boundary of the predefined range, this could be an indication that the parameter values might need to be re-calibrated after considering wider parameter ranges (Leta et al., 2015). However, some of the ranges cannot be further widened without violating physical constraints. The interactions between different hydrological processes may lead to poor parameter identifiability because some parameters compensate for others to make the simulated hydrograph match the observed one.

In the traditional approach (ExpTrad), all the parameters are treated equally within their *a prior* determined parameter ranges and are explored simultaneously using the same resolution, making them prone to the equifinality problem if there are a large number of parameters involved. This is because there are more chances that some of the parameters are affected by others. In addition, the presence of different sources of uncertainty, e.g., from the model structure, could add more complexity to the search thus further compounding the problem and generating the possibility of unrealistic parameter estimates. Taken together, the parameters calibrated with ExpTrad are thus less likely to be able to replicate the real behavior of the basin. In contrast, the calibration schemes, ExpD-U, ExpU-D, and ExpRand, based on the new framework, not only significantly reduce the search space but also have the search accomplished through several heterogeneous search subtasks. By reducing the chances of parameter interactions and many unnecessary parameter combinations, the results show that the new framework can mitigate the severe equifinality problem, partially overcome the effect of model structure error, and provide much more reliable parameter estimations.

Finally, Figure 7 shows the evolution of the best NSE obtained for the ten optimization trials. It can be seen that the three new framework-based configurations exhibit faster increase in the NSE objective function than the traditional approach of ExpTrad by the 950[th] model evaluation which corresponds to the moment when the first three sensitive parameter groups are calibrated (see Table 1). Table 8a shows that the NSE values of all three configurations (i.e., ExpD-U, ExpU-D, and ExpRand) are higher than those of ExpTrad in eight out of ten trials. The average NSEs of the ten trials are 0.74 for ExpD-U and ExpRand and 0.73 for ExpU-D while the average NSE for ExpTrad is 0.72 (see Table 8b). The NSE values of the three configurations stabilize after around 1500[th] model evaluations (see Figure 7), demonstrating rough convergence to a region with the highest probability. This corresponds to the moment when the first four groups of parameters are calibrated (see Table 1). By the 1500[th] model evaluation, the average NSEs of the ten trials are all 0.76 for the three new framework-based configurations while the average NSE for ExpTrad is 0.75. When reaching the 4000[th] evaluation, the average NSEs of the ten trials are 0.79 for ExpD-U and ExpRand and 0.78 for ExpU-D, with less than 7% increase in their NSEs from the 950[th] evaluation for all three configurations; while the average NSE of ExpTrad reaches 0.8 with more than 11% increase in its NSE from the 950[th] evaluation (see Table 8b). In addition, Figure 7 shows that ExpTrad has a number of "jumps" in its NSE values and its NSE evolution curves are the least smooth ones compared to those for the three configurations (i.e., ExpD-U, ExpU-D, and ExpRand) of the new framework in all ten panels.

Although each jump in ExpTrad leads to some improvements in its NSE, such improvements are unpredictable. The larger flexibility of the parameter ranges and the use of the same high resolution for all parameters in the search space simultaneously lead to higher chances of parameter interactions in ExpTrad, and they are also the sources causing such random jumps. The behavior of these jumps illustrates the randomness nature in ExpTrad's convergence. Furthermore, even though ExpTrad has the highest average NSE value (0.8) versus an average of NSE = 0.79 for ExpD-U and ExpRand, and an average of NSE = 0.78 for ExpU-D (see Table 8b), the estimated parameter values by ExpTrad are significantly less reliable than those by our new framework for all three configurations as measured by NC and HB (see Tables 6 and 7). This demonstrates that the traditional calibration approach, such as ExpTrad, could yield less errors (i.e., better NSEs) for wrong reasons (i.e., unrealistic parameter values) and that exemplifies the serious parameter equifinality problem we face today. The ExpTrad's fixed parameter ranges and its strategy of varying all parameters simultaneously using the same resolution are the sources that lead to the various unrealistic parameter combinations in the parameter space. This equifinality problem becomes more severe with the increase of the number of parameters to be calibrated.

In contrast, all three of the new framework-based calibrations lead to smoother (i.e., gradual but mostly constant improvement) and more robust convergence behavior as demonstrated in all ten trials in Figure 7. In addition, Figure 7 shows that a majority of our new framework usually has better performance at the beginning of the optimization process where the NSE evolution curves are more often than not above those of the traditional method in the first 1500 iterations, suggesting that the continuous and gradual progress achieved by the new framework also leads to faster NSE improvements. Moreover, the parameter values obtained with all of the three configurations for all ten trials are more reasonable than those with the widely used traditional calibration approach (Tables 6 and 7). These superior performances by the new framework indicate that our framework has an effective strategy in searching for more realistic parameter combinations in the parameter space and the search is efficient. Also, the smoother NSE evolution curves indicate that the proposed framework at least guarantees suboptimal, if not optimal, parameter estimations in terms of its objective function, such as NSE.

As for the application of the proposed new framework, among the three configurations, ExpD-U has the lowest number of NC parameters and least number of parameters hitting the boundary, and produces the largest number of parameters that are close to their "true" values. We attribute the superior performance of ExpD-U over the other two configurations of ExpU-D and ExpRand to the prioritization of the soil-related parameters of downstream cells which, given their proximity to the outlet, have a larger effect on the rising limbs and the peaks of the storm hydrographs. We expect this to be the case for watersheds that have relatively homogeneous spatial distribution of rainfalls provided the streamflow at the outlet is used for calibration.

In summary, results of this 134 parameter calibration study (i.e., Figures 6 and 7 and Tables 6 and 7) clearly demonstrate that the use of the principle of multiresolution based on parameters' importance, the constant adjustment of the parameters' variation ranges, and the heterogeneous calibration procedures with varying number of execution runs can dramatically reduce the equifinality problem. This is clearly reflected in the more realistic parameter estimates obtained, the better consistency toward convergence (smoother NSE evolution curves), and the faster improvement attained. For complex hydrological models with a large number of parameters, we have shown that our new calibration framework is effective and capable.

## 5. Conclusions

This study introduces a new calibration framework for high-resolution hydrological models that are associated with a large number of parameters. This new framework combines the multiresolution principle with heterogeneous calibration procedures. Our framework divides the calibration problem into several sub-tasks or runs of the optimization algorithm in which different parameter groups are searched with different resolutions (e.g., see Table 1) according to their importance (i.e., multiresolution principle). Across the runs, the targeted groups are shifted and the search ranges and resolutions for each group of parameters are adjusted (i.e., heterogeneous calibration procedure).

Three different configurations were tested in a quasi-synthetic experiment with a double-model setup using two well-known modeling engines: VIC and DHSVM. The VIC model for the French Creek watershed in Pennsylvania was first calibrated to match the observed streamflow at the outlet. Then the VIC model with 21 soil textures was used to generate a synthetic ground truth. The DHSVM model with 134 parameters was calibrated to match the synthetic truth from VIC. This double-model setup manages to provide much more control over the "ground truth" compared to using direct measurements and, at the same time, incorporates the structural uncertainties of models that are neglected in fully-synthetic experimental setups found in previous studies in the literature.

The tests compared three configurations using the new framework and a traditional optimization-based calibration approach with ten trials, or 40,000 model evaluations. Four metrics were used on the resulting solution sets: (1) the ratio of parameters that converges; (2) the ratio of parameters that hits the boundary of the allowed ranges; (3) the differences between the optimized values and the available "true" values; and (4) the evolution of the Nash-Sutcliffe efficiency (NSE) coefficient. Our test results from these ten trials show that the new calibration framework leads to higher ratios of converged and non-boundary parameter values, smaller differences with the synthetic reference parameter values, a more robust and smoother improvement in terms of NSE values, and usually an initially faster rate of improvement than the traditional calibration approach.

Even though our proposed framework produced slightly lower average NSE values, i.e., 0.79 for ExpD-U and ExpRand and 0.78 for ExpU-D versus 0.80 for ExpTrad, the new framework resulted in at least comparable NSE values. Moreover, due to its gradual reduction of the uncertainty, the new framework produced smoother convergence trends of the NSE values than the traditional method which exhibited many random jumps (see Figure 7). Similarly, the accelerated rate of improvement could be translated into higher computational efficiency in computation-constrained environments. These results attest the advantages of the new framework in tackling calibration problems of high dimensionality because of its ability to mitigate the problems related to equifinality with more realistic parameter estimations, and because of its potential to achieve results of competent quality when smaller number of model evaluations is available (see Figure 7 and Table 8). These promising characteristics of the new framework are attributed to the use of the multiresolution principle in the parameter search space with heterogeneous calibration procedures, which makes it possible to discern among a large number of parameter combinations, since the resolution is adjusted and the population size is reduced in the search space in each run or sub-task (see Table 1 and Figure 2).

In our synthetic experiment case study, we calibrated 134 hydrological model parameters. However, the proposed framework is versatile and is applicable for other types of distributed models particularly when there are a large number of model parameters to be calibrated. In the future, multiple avenues of improvements for our framework will be explored. First, an extensive study on impacts of different running plans, such as the selection of the number of model evaluations, $n(r)$, for each run will be thoroughly investigated. Also, in order to have a relatively balanced number of parameters in each group, we would like to further investigate how to effectively prioritize parameters to different groups to mitigate situations where an overwhelmingly large number of parameters have similar sensitivities and how such prioritization strategy would affect the framework's performance. In addition, we would like to investigate how the use of different observations (e.g., streamflow, soil moisture, and evapotranspiration) may affect the prioritization strategies in grouping the parameters to achieve a relatively balanced size for each group. For example, in this study, the streamflow at the outlet is used. But if the spatial distribution of soil moisture rather than streamflow at the outlet is used with the same running plan, the preference of the three configurations, i.e., ExpD-U, ExpU-D, and ExpRand, may change, implying that the prioritization strategy for the ranks of the relevant groups may be different. In addition, we would like to investigate how sensitive the number of discretizations in the lower ranked groups may have on the overall results. Second, we would like to investigate the performance of the framework in combination with different types of sensitivity analysis methods to explore the impacts of granularity and complexity of the different sensitivity analysis methods on the effectiveness of our framework. Third, we would like to investigate the performance of our framework in combination with different types of global optimization algorithms, and also to extend the calibration framework to optimize multiple objectives simultaneously (e.g., streamflow and spatially distributed soil moisture) or to use a formal Bayesian approach (Schoups & Vrugt, 2010; Sun et al., 2017) which might help further reducing the uncertainty in high-resolution models. Fourth, we would like to apply our framework to real world cases when detailed spatial observations become available to thoroughly explore the strengths and weaknesses of our new framework. Finally, some studies have pointed out that parameters associated with subsurface fluxes optimized poorly in the traditional model calibration approaches as they rely on observed streamflow (Immerzeel & Droogers, 2008; Rajib et al., 2016; Wanders et al., 2014). We thus intend to test if calibrations using our new framework could better estimate parameters related to subsurface fluxes.

## Acknowledgments

This work was supported in part by the United States Department of Transportation through award no. OASRTRS-14-H-PIT to the University of Pittsburgh and by the William Kepler Whiteford Professorship to the third author from the University of Pittsburgh. The first author would like to thank the China Scholarship Council for sponsoring him to pursue his PhD research under the supervision of Professor Xu Liang at the University of Pittsburgh. The fourth author acknowledges the partial support of the National Natural Science Foundation of China (41675109). We also thank the Center for Research Computing and the Swanson School of Engineering at the University of Pittsburgh, the High-Performance Computing Center of Nanjing University, and the China Postdoctoral Science Foundation (2019M661714) for providing access to their computing resources.

For this work, Ruochen Sun and Felipe Hernández implemented the research ideas, performed the experiment, conducted the analysis, and co-wrote the initial draft. Xu Liang conceived the research ideas, designed the experiment, supervised the investigation, and co-wrote and finalized the manuscript. Huiling Yuan contributed in refining the NC and HB evaluation metrics.

All the data utilized for the VIC (https://vic.readthedocs.io/en/master/) and DHSVM (https://dhsvm.pnnl.gov/) models are publicly available from the U.S. government agency websites, USGS streamflow data (https://waterdata.usgs.gov/nwis/sw) and NASA NLDAS-2 forcing data (https://disc.gsfc.nasa.gov/datasets?keywords=NLDAS). All the data generated in this work for the figures and tables will be available through Mendeley Data (DOI:

10.17632/68p8d76rwh.2). The VIC, DHSVM, and optimization algorithm (https://github.com/felherc/MAESTRO_MO) used in this work are all open sources.

**Figure captions:**

Figure 1. Flowchart of the new calibration methodology

Figure 2. Example convergence for three parameters throughout a four-run calibration process. The blue bars represent the allowable continuous search range in the parameter space that is associated with a fine resolution, which shrinks after each run; green markers indicate allowable discrete assignments in the parameter search space that are associated with coarser resolutions. The orange bars in the last row (i.e., Result) represent the range of variation of the three parameters in the final population.

Figure 3. Map of the French creek watershed displaying the water flow time to the outlet and the 21 VIC model cells with 1/32$^{th}$ degree resolution. The symbols in each cell represent a division of the watershed based on flow time. "D" denotes the downstream region; "M" denotes the midstream region; "U" denotes the upstream region.

Figure 4. Comparison of the daily streamflow time series simulated by the VIC model with the USGS observations. The calibration period is from 2004 to 2011, with 2003 used for spin-up.

Figure 5. (a) The soil texture triangle defined by the USDA. 21 soil textures (white dots) are selected to design the synthetic experiment. (b) Soil texture map of the watershed with resolution of the DHSVM model (500m). The ID numbers correspond to the 21 soil textures which are assigned to the 21 VIC model cells.

Figure 6. Tukey simultaneous 95% confidence intervals of differences of means for (a) NC and (b) HB. If one interval does not contain zero, the corresponding means are significantly different.

Figure 7. Evolution of the Nash–Sutcliffe Efficiency (NSE) coefficient for the ten optimization trials.

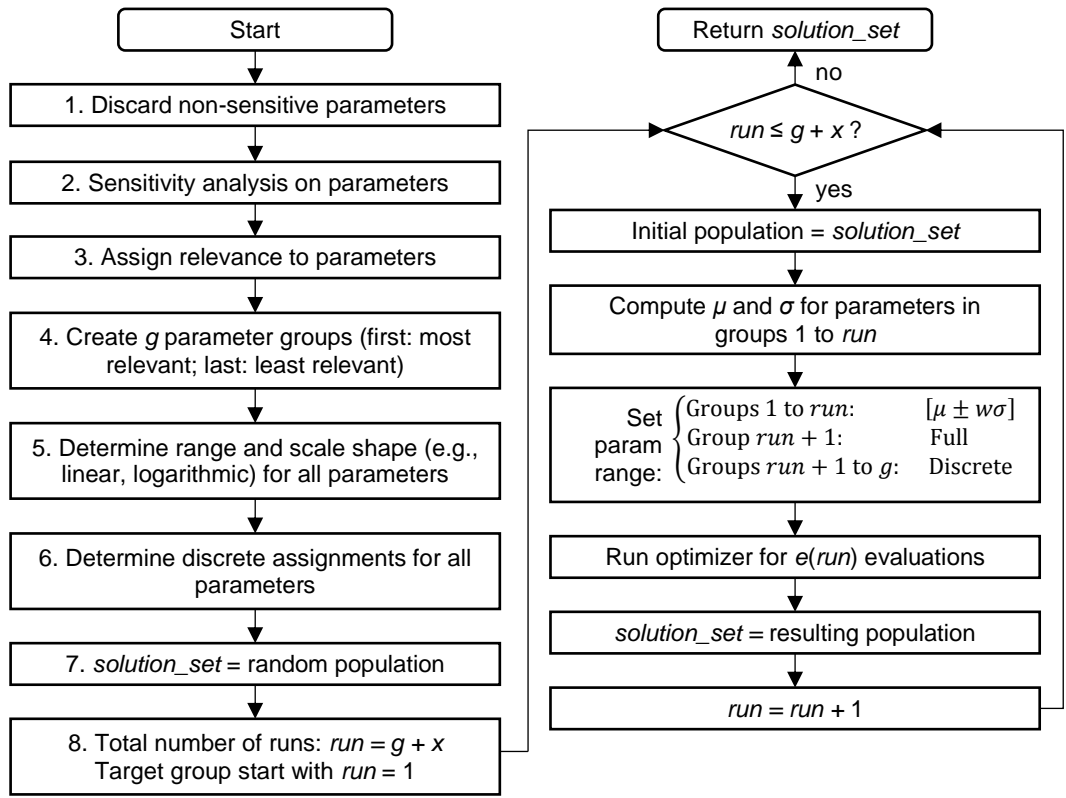

Figure 1. Flowchart of the new calibration methodology

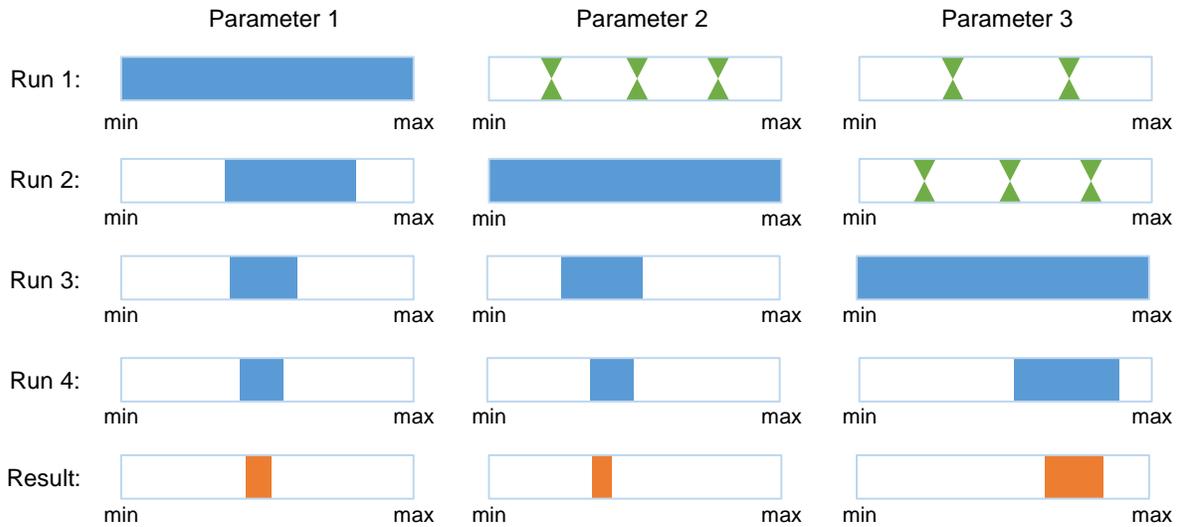

Figure 2. Example convergence for three parameters throughout a four-run calibration process. The blue bars represent the allowable continuous search range in the parameter space that is associated with a fine resolution, which shrinks after each run; green markers indicate allowable discrete assignments in the parameter search space that are associated with coarser resolutions. The orange bars in the last row (i.e., Result) represent the range of variation of the three parameters in the final population.

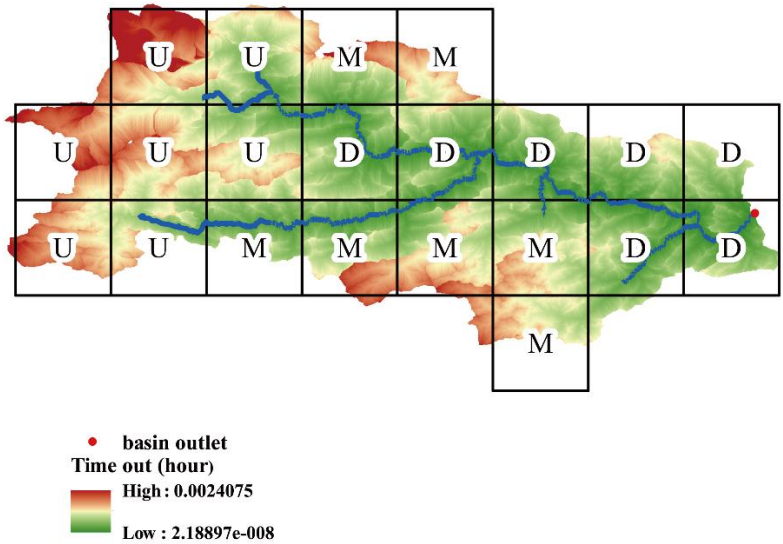

Figure 3. Map of the French creek watershed displaying the water flow time to the outlet and the 21 VIC model cells with 1/32th degree resolution. The symbols in each cell represent a division of the watershed based on flow time. "D" denotes the downstream region; "M" denotes the midstream region; "U" denotes the upstream region.

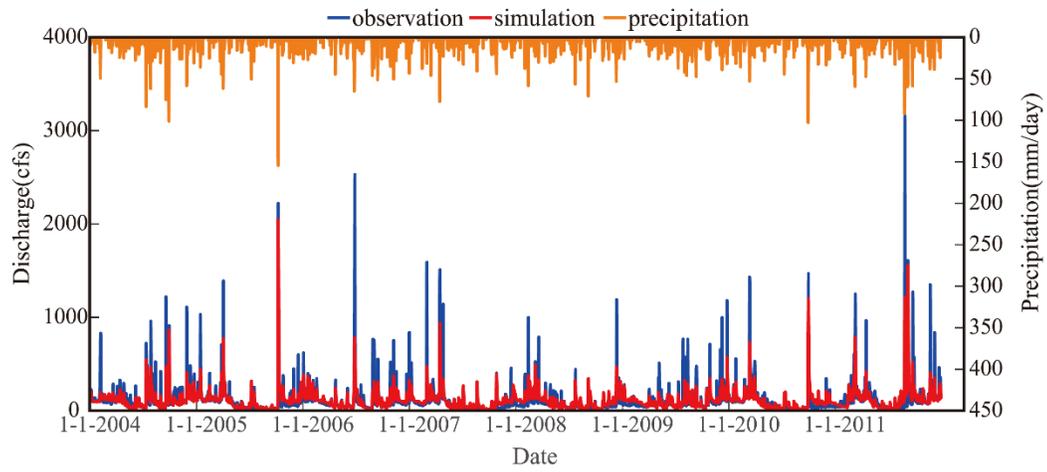

Figure 4. Comparison of the daily streamflow time series simulated by the VIC model with the USGS observations. The calibration period is from 2004 to 2011, with 2003 used for spin-up.

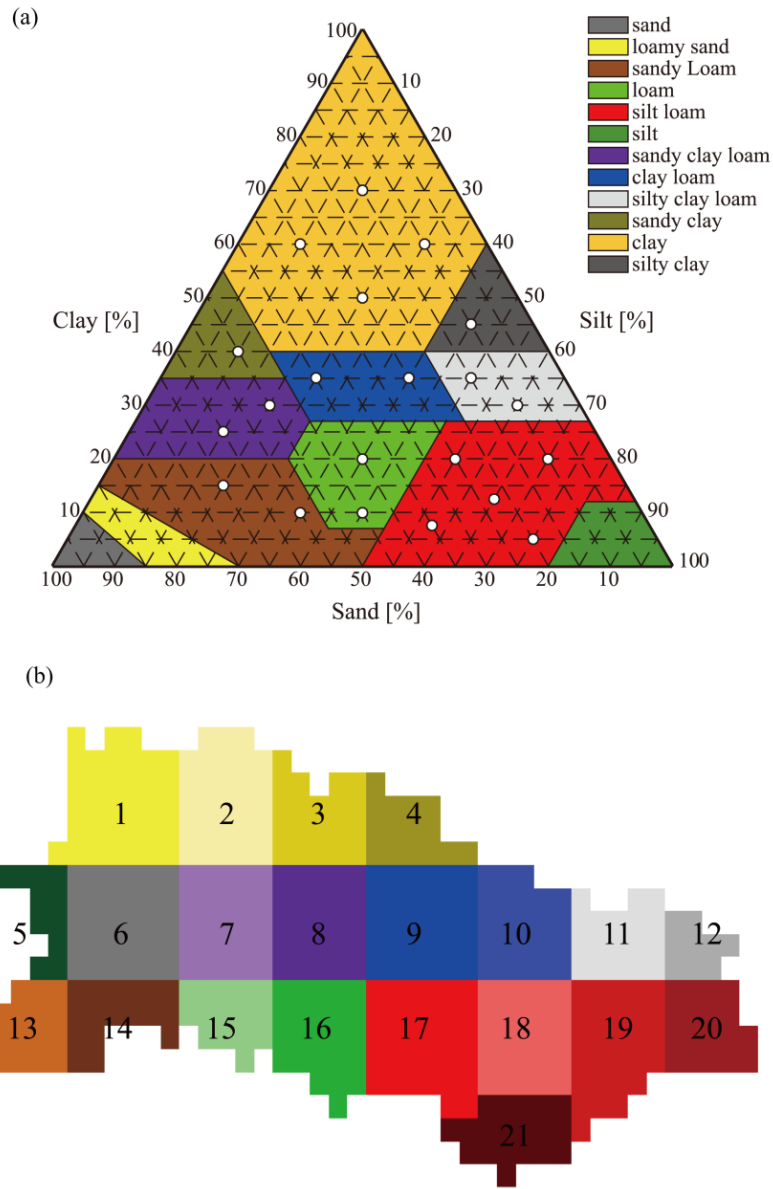

Figure 5. (a) The soil texture triangle defined by the USDA. 21 soil textures (white dots) are selected to design the synthetic experiment. (b) Soil texture map of the watershed with resolution of the DHSVM model (500m). The ID numbers correspond to the 21 soil textures which are assigned to the 21 VIC model cells.

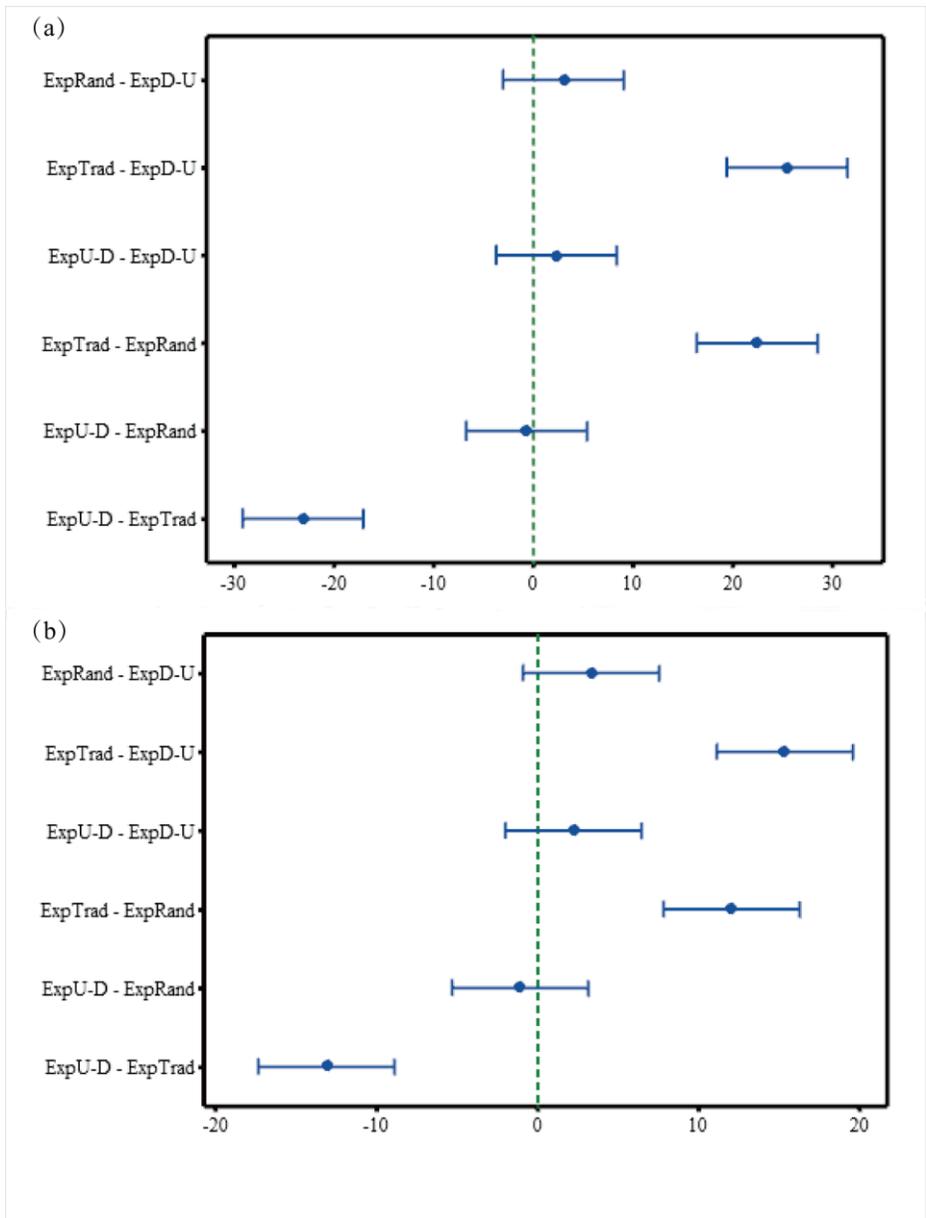

Figure 6. Tukey simultaneous 95% confidence intervals of differences of means for (a) NC and (b) HB. If one interval does not contain zero, the corresponding means are significantly different.

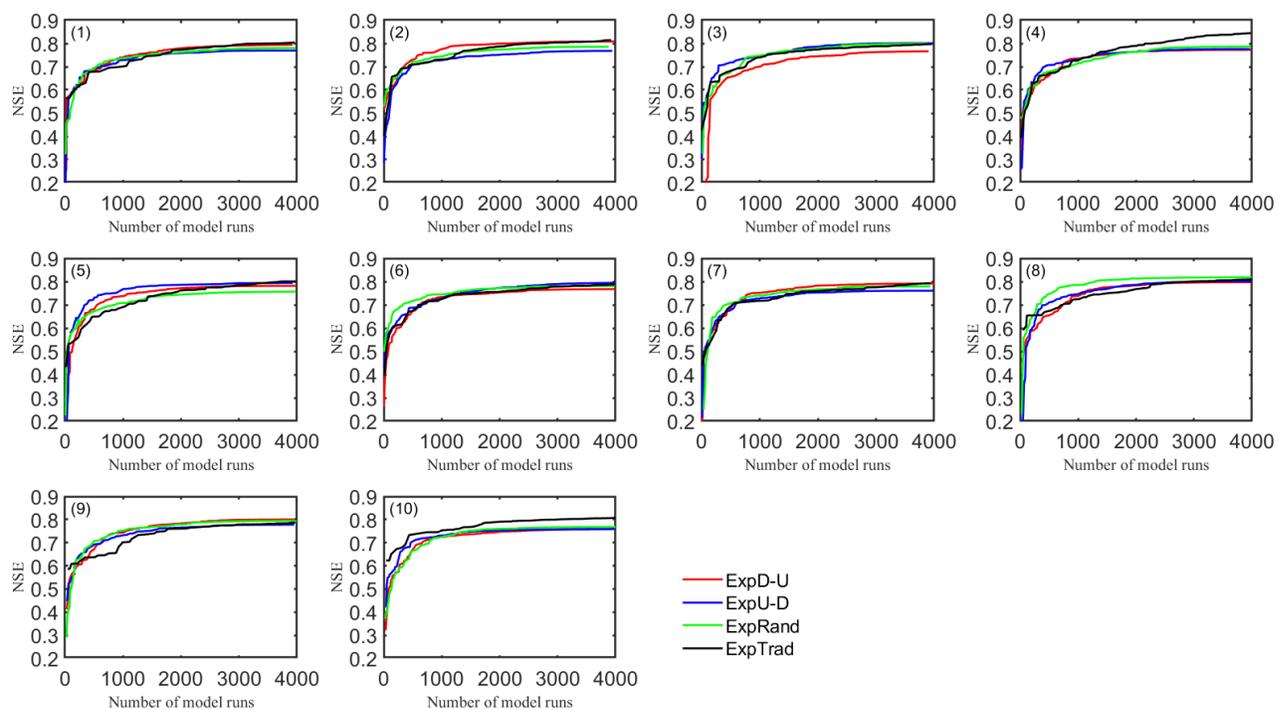

Figure 7. Evolution of the Nash–Sutcliffe Efficiency (NSE) coefficient for the ten optimization trials.

Table 1

*Setup of the calibration schemes based on the new calibration framework*

|  | Run 1 | Run 2 | Run 3 | Run 4 | Run 5 | Run 6 | Run 7 |
|---|---|---|---|---|---|---|---|
| *Number of model executing/running times* | | | | | | | |
|  | 200 | 300 | 450 | 550 | 700 | 800 | 1000 |
| *Evolution of parameter ranges[a] and number of parameter discrete values* | | | | | | | |
| Group1 | full | shrunk | shrunk | shrunk | shrunk | shrunk | shrunk |
| Group2 | 5[b] | full | shrunk | shrunk | shrunk | shrunk | shrunk |
| Group3 | 5 | 5 | full | shrunk | shrunk | shrunk | shrunk |
| Group4 | 5 | 5 | 5 | full | shrunk | shrunk | shrunk |
| Group5 | 5 | 5 | 5 | 5 | full | shrunk | shrunk |
| Group6 | 2 | 2 | 2 | 3 | 5 | full | shrunk |

[a]As described in section 2, "full" means parameters are calibrated with their original full ranges, while "shrunk" means parameters are calibrated with their updated ranges that have been shrunk after the last run. The meta-parameter $w$ is set to 2.

[b]The integer "5' means that parameters in Group 2 are calibrated using five discrete values within their variation ranges.

Table 2

*A list of conceptual parameters for the VIC model*

| Parameters | Meaning | Typical range | Units | Calibrated value |
|---|---|---|---|---|
| $b$ | Exponent of variable infiltration capacity curve | 0-0.4 | – | 0.36 |
| $Ws$ | Fraction of maximum soil moisture content of the lowest layer where non-linear baseflow occurs | 0-1.0 | – | 0.99 |
| $Dsmax$ | Maximum velocity of base flow | 0-30.0 | mm/d | 1.63 |
| $Ds$ | Fraction of $Dsmax$ where non-linear baseflow begins | 0-1.0 | – | 0.88 |
| $d2$ | The depth of the second soil layer | 0.1-2.0 | m | 0.17 |
| $d3$ | The depth of the third soil layer | 0.1-2.0 | m | 0.11 |

Table 3

*A list of soil parameters in the DHSVM Model corresponding to the 21 soil textures*

| Soil texture ID | Soil type[a] | Field capacity | | Porosity | | Wilting point | | Vertical saturated hydraulic conductivity (m/s) | |
|---|---|---|---|---|---|---|---|---|---|
| | | Range | "True" value | Range | "True" value | Range | "True" value | Range | "True" value |
| 1 | c | 0.15-0.6 | 0.54 | 0.1-0.65 | 0.62 | 0.1-0.55 | 0.41 | 1E-8-1E-5 | 7.62E-7 |
| 2 | c | 0.15-0.6 | 0.45 | 0.1-0.65 | 0.56 | 0.1-0.55 | 0.34 | 1E-8-1E-5 | 4.14E-7 |
| 3 | c | 0.15-0.6 | 0.50 | 0.1-0.65 | 0.60 | 0.1-0.55 | 0.36 | 1E-8-1E-5 | 6.67E-7 |
| 4 | c | 0.15-0.6 | 0.42 | 0.1-0.65 | 0.54 | 0.1-0.55 | 0.28 | 1E-8-1E-5 | 4.80E-7 |
| 5 | sc | 0.15-0.5 | 0.32 | 0.25-0.55 | 0.50 | 0.12-0.35 | 0.22 | 1E-8-1E-5 | 4.14E-7 |
| 6 | sic | 0.15-0.55 | 0.42 | 0.15-0.6 | 0.56 | 0.1-0.4 | 0.26 | 1E-8-1E-5 | 7.70E-7 |
| 7 | scl | 0.1-0.45 | 0.25 | 0.2-0.6 | 0.47 | 0.07-0.2 | 0.15 | 1E-8-1E-5 | 1.17E-6 |
| 8 | scl | 0.1-0.45 | 0.28 | 0.2-0.6 | 0.48 | 0.07-0.2 | 0.17 | 1E-8-1E-5 | 8.11E-7 |
| 9 | cl | 0.1-0.6 | 0.32 | 0.13-0.65 | 0.50 | 0.07-0.35 | 0.20 | 1E-8-1E-5 | 6.67E-7 |
| 10 | cl | 0.1-0.6 | 0.35 | 0.13-0.65 | 0.52 | 0.07-0.35 | 0.19 | 1E-8-1E-5 | 8.69E-7 |
| 11 | sicl | 0.13-0.55 | 0.36 | 0.2-0.65 | 0.53 | 0.09-0.4 | 0.19 | 1E-7-1E-4 | 1.07E-6 |
| 12 | sicl | 0.13-0.55 | 0.34 | 0.2-0.65 | 0.53 | 0.09-0.4 | 0.17 | 1E-7-1E-4 | 1.64E-6 |
| 13 | sl | 0.05-0.3 | 0.21 | 0.13-0.6 | 0.44 | 0.02-0.15 | 0.11 | 1E-7-1E-4 | 3.69E-6 |
| 14 | sl | 0.05-0.3 | 0.22 | 0.13-0.6 | 0.41 | 0.02-0.15 | 0.09 | 1E-7-1E-4 | 7.33E-6 |
| 15 | l | 0.07-0.45 | 0.26 | 0.12-0.7 | 0.47 | 0.04-0.2 | 0.13 | 1E-7-1E-4 | 2.42E-6 |
| 16 | l | 0.07-0.45 | 0.24 | 0.12-0.7 | 0.43 | 0.04-0.2 | 0.09 | 1E-7-1E-4 | 7.42E-6 |
| 17 | sil | 0.05-0.5 | 0.31 | 0.2-0.7 | 0.49 | 0.03-0.3 | 0.12 | 1E-7-1E-3 | 3.69E-6 |
| 18 | sil | 0.05-0.5 | 0.28 | 0.2-0.7 | 0.45 | 0.03-0.3 | 0.10 | 1E-7-1E-3 | 6.50E-6 |
| 19 | sil | 0.05-0.5 | 0.29 | 0.2-0.7 | 0.48 | 0.03-0.3 | 0.12 | 1E-7-1E-3 | 2.92E-6 |
| 20 | sil | 0.05-0.5 | 0.26 | 0.2-0.7 | 0.41 | 0.03-0.3 | 0.09 | 1E-7-1E-3 | 9.89E-6 |
| 21 | sil | 0.05-0.5 | 0.30 | 0.2-0.7 | 0.38 | 0.03-0.3 | 0.09 | 1E-7-1E-3 | 1.16E-5 |

[a]The symbols of soil type in this table are acronyms of soil texture classifications defined by the USDA. The classifications appeared in the table include clay (c), sandy clay (sc), silty clay (sic), sandy clay loam (scl), clay loam (cl), silty clay loam (sicl), sandy loam (sl), loam (l), and silt loam (sil).

Table 4

*A List of additional DHSVM Parameters which need to be calibrated*

| Parameters | Units | Range | Referenced "true" value from VIC |
|---|---|---|---|
| *Soil parameters* | | | |
| Lateral saturated hydraulic conductivity | m/s | 1.0E-5-0.1 | - |
| Exponent for change of lateral conductivity with depth (Exponential decrease) | - | 0-10 | - |
| *Vegetation parameters* | | | |
| Minimum stomatal resistance for the overstory | s/m | 0-200 | 80 |
| Minimum stomatal resistance for the understory | s/m | 0-200 | 80 |
| Radiation attenuation | - | 0.1−0.8 | 0.5 |
| Fractional coverage of overstory | - | 0−1 | - |
| Aerodynamic attenuation | - | 0−3 | - |
| Soil moisture threshold to restrict transpiration for the overstory | - | 0−1 | - |
| Soil moisture threshold to restrict transpiration for the understory | - | 0−1 | - |
| *Routing parameter* | | | |
| Manning's coefficient of the river channels | - | 1.0E-6−0.1 | - |

Table 5

*A List of groups of the 134 DHSVM model parameters for the synthetic experiments with the French Creek watershed*

| Group | Parameters | Parameter numbers |
|-------|------------|-------------------|
| 1 | field capacity (×7); lateral saturated hydraulic conductivity (×7); exponential decrease (×7); Manning's coefficient | 22 |
| 2 | field capacity (×7); lateral saturated hydraulic conductivity (×7); exponential decrease (×7) | 21 |
| 3 | field capacity (×7); lateral saturated hydraulic conductivity (×7); exponential decrease (×7) | 21 |
| 4 | porosity (×10); wilting point (×10); minimum stomatal resistance for overstory; minimum stomatal resistance for understory; fractional coverage of overstory | 23 |
| 5 | porosity (×11); wilting point (×11) | 22 |
| 6 | vertical saturated hydraulic conductivity (×21); radiation attenuation; aerodynamic attenuation; moisture threshold for overstory; moisture threshold for understory | 25 |

Table 6

*Comparison of the different calibration schemes based on three evaluation metrics for the synthetic experiments with the French Creek watershed*

|  | ExpD-U | ExpU-D | ExpRand | ExpTrad |
|---|---|---|---|---|
| NC | 5.8 | 8.2 | 8.9 | 31.3 |
| HB | 8.5 | 10.7 | 11.8 | 23.8 |
| RosARE | 54.8% | 50.2% | 50.7% | - |

Note: Only ExpD-U produced a mean RosARE that was statistically significantly higher than 50% with p-value = 0.016, while the p-values for ExpU-D and ExpRand are greater than 0.05.

Table 7

*Number of not converged parameters within each parameter group for the synthetic experiments with the French Creek watershed*

|  | Group1 | Group2 | Group3 | Group4 | Group5 | Group6 |
|---|---|---|---|---|---|---|
| *Grouping based on ExpD-U* | | | | | | |
| ExpD-U |  |  |  | 0.6 |  | 5.2 |
| ExpTrad | 2.9 | 3.5 | 3.4 | 8 | 5.1 | 8.4 |
| *Grouping based on ExpU-D* | | | | | | |
| ExpU-D |  |  | 0.3 | 0.3 | 2 | 5.6 |
| ExpTrad | 2.3 | 3.5 | 4 | 5.1 | 8 | 8.4 |
| *Grouping based on ExpRand* | | | | | | |
| ExpRand |  |  | 0.2 | 0.4 | 1.3 | 7 |
| ExpTrad | 1.9 | 3.9 | 3.9 | 5.3 | 8 | 8.3 |

Table 8a.

*A summary of NSE values for each configuration for 10 trials, where X represents the number of model evaluations, with the French Creek watershed*

|    | ExpD-U | | ExpU-D | | ExpRand | | ExpTrad | |
|----|-----|-------|-----|-------|-----|-------|-----|-------|
|    | X   | NSCE  | X   | NSCE  | X   | NSCE  | X   | NSCE  |
| 1  | 950 | 0.739 | 950 | 0.727 | 950 | 0.736 | 950 | 0.699 |
| 2  |     | 0.774 |     | 0.726 |     | 0.742 |     | 0.729 |
| 3  |     | 0.698 |     | 0.741 |     | 0.746 |     | 0.738 |
| 4  |     | 0.734 |     | 0.729 |     | 0.710 |     | 0.726 |
| 5  |     | 0.736 |     | 0.760 |     | 0.707 |     | 0.687 |
| 6  |     | 0.732 |     | 0.723 |     | 0.745 |     | 0.720 |
| 7  |     | 0.750 |     | 0.726 |     | 0.739 |     | 0.716 |
| 8  |     | 0.737 |     | 0.744 |     | 0.785 |     | 0.723 |
| 9  |     | 0.743 |     | 0.729 |     | 0.751 |     | 0.693 |
| 10 |     | 0.724 |     | 0.730 |     | 0.722 |     | 0.747 |

Table 8b.

*A summary of averaged NSE values of 10 trials for each configuration, where X represents the number of model evaluations, with the French Creek watershed*

|         | ExpD-U | | ExpU-D | | ExpRand | | ExpTrad | |
|---------|------|------|------|------|------|------|------|------|
|         | X    | NSE  | X    | NSE  | X    | NSE  | X    | NSE  |
| Average | 950  | 0.74 | 9500 | 0.73 | 950  | 0.74 | 950  | 0.72 |
| Average | 1500 | 0.76 | 1500 | 0.76 | 1500 | 0.76 | 1500 | 0.75 |
| Average | 4000 | 0.79 | 4000 | 0.78 | 4000 | 0.79 | 4000 | 0.80 |